\documentclass{appolb}
\usepackage{epsfig}
\usepackage{amsmath}
\usepackage{colordvi,graphicx}
\usepackage{graphicx}  	
\usepackage{mathrsfs}

\begin{document}


\title{Real space Bogolubov-de Gennes equations study\\
of the boson-fermion model
}
\author{Jaromir Krzyszczak, Tadeusz Doma\'nski, Karol I. Wysoki\'nski
\address{Institute of Physics and Nanotechnology Centre,\\ 
       M.\ Curie - Sk\l odowska University, 20-031 Lublin, Poland}
}
\maketitle


\begin{abstract}

The boson-fermion (BF) model has been 
proposed to describe  superconductivity in short coherence
length superconductors. 
In this work we use it to study 
impurity induced inhomogeneities in high 
temperature superconductors (HTS) as found in
 numerous scanning tunnelling measurements. 
The model has been formulated in real space 
and solved with help of Bogolubov-de Gennes 
approach. Disorder in the boson or fermion 
subsystem is directly coupled to the superconducting 
order parameter and leads to severe changes of 
superconducting properties like local order parameter 
and density of states. We present the results 
for many impurities randomly distributed over 
otherwise clean and periodic 
two-dimensional square lattice.
\end{abstract}


\section{Introduction}
The boson-fermion model assumes the existence in the 
system of two kinds of interacting quantum objects \cite{Ranning85} 
 consisting of (usually) local and immobile bosonic 
particles  (doubly charged preformed pairs) 
interacting with mobile fermions (electrons).
It is the boson-fermion scattering which induces the condensation
transition in bosonic  and superconducting transition 
in fermionic subsystem \cite{Kostyrko}. 
In real materials, as e.g. doped HTS, one expects
both bosonic and fermionic parameters to depend on the 
local environment and thus to be random variables. 

Here we present results of calculations of such
important parameters as local value of the gap parameter
$\Delta_i$  and local density of 
states $N_i(E)$ at lattice
site $i$. $\Delta_i$ and $N_i(E)$ are subject to 
measurements with help of scanning tunnelling microscope (STM).
The STM spectra of many high temperature superconductors \cite{STM},
{\it e.g.} Bi$_2$Sr$_2$CaCu$_2$O$_{8+\delta}$, 
indeed show the intrinsic inhomogeneities of $\Delta_i$ and $N_i(E)$
extending over the scales of few nanometres. 

Even though the microscopic origin of these inhomogeneities 
remains unclear  some experiments show positive correlation between
location of the oxygen dopant and the value of local order parameters.
Based on this scenario theoretical approaches
have been proposed   in which authors calculate 
local properties of the system by means of real space 
Bogolubov-de Gennes approach, using Hubbard \cite{Nunner05},
t-J \cite{Maska07} or similar models with disorder in 
normal (density) or anomalous (Cooper) channel \cite{Anderson06}.
To account for possible sources of inhomogeneities
we use the boson-fermion model \cite{Ranning85} in the form
\begin{eqnarray}
\hat{H}^{\text{BF}} &=&
\sum_{i,j,\sigma} t_{i j} \hat{c}_{i\sigma}^{\dagger} \hat{c}_{j\sigma} +
 \sum_{i\sigma} \left( V_{i}^{\text{imp}} -
 \mu \right) \hat{c}_{i\sigma}^{\dagger} \hat{c}_{i\sigma} + \\
\nonumber
&+& \sum_{i} \left( E^{B} + \delta E_{i}^{B} 
- 2\mu \right) \hat{b}_{i}^{\dagger} \hat{b}_{i}
+ \sum_{i,j} \frac{g_{i j}}{2} \left( \hat{b}_{i}^{\dagger} 
\hat{c}_{i\downarrow} \hat{c}_{j\uparrow} + \hat{b}_{i} 
\hat{c}_{i\uparrow}^{\dagger} \hat{c}_{j\downarrow}^{\dagger} \right),
\label{BF_hamilt}
\end{eqnarray}
where   $i$ and $j$ denote lattice sites of the  square lattice, 
$\hat{c}_{i,\sigma}^{\dagger}$ ($\hat{c}_{i,\sigma}$) 
stand for creation (annihilation) operator  
of fermion at the site $i$ with spin $\sigma$ and
 $\hat{b}_{i}^{\dagger}$ and $\hat{b}_{i}$ are creation
 and annihilation operators of hard-core bosons at the site $i$.
$\mu$ denotes chemical potential of the system and $t_{i j}$ 
are hopping integrals. For numerical calculations we assume 
 hopping integrals to nearest and next nearest 
neighbours $t_1, t_2$ to be
different from zero. As noted earlier the presence of oxygen impurities in Bi family of 
HTS on one hand is the source of free carriers (holes) 
 and at the same time introduces disorder 
into the materials. Undoped system is insulating.
 We assume that this disorder can be modelled
by  random on-site energies $V_{i}^{\text{imp}}$ in fermion subsystem 
and random hard-core boson energies 
$\delta E_{i}^{B}$ \cite{Robaszk01,Doman02}. 
To account for the short coherence length and
d-wave symmetry of the superconducting order
parameter we assume that the boson-fermion coupling $g_{i j}$ 
takes on non-zero value  for nearest neighbour 
sites $<i,j>$ only and is equal +g if $j=i\pm\vec{x}$ 
and -g if $j=i\pm\vec{y}$.

Application of standard Hartree-Fock-Bogolubov decoupling 
leads to $\hat{H}^{\text{BF}}=
\hat{H}^{\text{B}}+\hat{H}^{\text{F}}$, where 
$\hat{H}^{\text{B}}$ is single site bosonic hamiltonian with parameters
depending on the fermionic order parameter.  
$\hat{H}^{\text{F}}$ is the Hamiltonian describing fermionic subsystem \cite{Doman02}. 
Standard statistical approach allows the exact 
solution of the boson Hamiltonian $\hat{H}^{\text{B}}$. One finds  
\begin{eqnarray}
<\hat{b}_{i}^{\dagger} \hat{b}_{i} > &=& \frac{1}{2} - \frac{E^{B} 
+ \delta E_{i}^{B} - 2\mu}{4\gamma_{i}} \tanh \left( \frac{\gamma_{i}}{k_{B}T} \right) \\
<\hat{b}_{i}> &=& - \frac{\chi_{i}}{2\gamma_{i}} \tanh \left( \frac{\gamma_{i}}{k_{B}T} \right).
\label{ord_pars}
\end{eqnarray}
Here $\gamma_{i} = \sqrt{ \left( \frac{E^{B} + \delta E_{i}^{B} 
- 2\mu}{2} \right)^{2} + |\chi_{i}|^{2}}$,  $\chi_{i} = \displaystyle \sum_{<j>} \frac{g_{i j}}{2}
 <\hat{c}_{i,\downarrow} \hat{c}_{j,\uparrow}>$ and $\Delta_{ij} 
= \frac{g_{j i}}{2} <\hat{b}_{j}>$. It is $\Delta_{ij}$ which couples two subsystems
as is evident from equation (\ref{eq_BdG}) below.

The fermion part has the standard BCS-structure 
and we diagonalise it by the Bogolubov-Valatin transformation, 
which yields the following Bogolubov-de Gennes equations \cite{Doman02}  
\begin{eqnarray}
\sum_{j}
\left(
\begin{array}{c}
t_{i j} + \left( V_{i}^{\text{imp}} - \mu \right) \delta_{ij};  \Delta_{ij} \\
\Delta_{ij}^{*};
 - t_{i j} - \left( V_{i}^{\text{imp}} - \mu \right) \delta_{ij}
\end{array}
\right)
\left(
\begin{array}{r}
u^{l}_{j} \\
v^{l}_{j}
\end{array}
\right)=
E^{l} \left(
\begin{array}{c}
u^{l}_{i} \\
v^{l}_{i}
\end{array}
\right)
\label{eq_BdG}
\end{eqnarray}

We solve numerically the above Bogolubov-deGennes equations 
in real space iteratively to get 
self-consistent values for all $\Delta_{ij}$ with high accuracy.
Even though our model and calculations are motivated 
by the experimental results obtained mainly for Bi 
family high temperature superconductors we do not 
try to model the actual experimental situation here. Instead we 
shall present the results on local properties of
disordered boson-fermion model. 
In this respect the paper
is an extension of the previous 
CPA calculations \cite{Doman02} of the
same model.
\section{Results and discussion}
Before we start the discussion of the results
let us remind that it is the position of
the boson level with respect to
the Fermi level $E_B$ which tells us if the system will be
 superconducting at all. Due to phase space
restrictions at low temperatures 
the scattering of preformed bosons
is allowed  only, if the bosonic level is close enough 
to the Fermi level.

To account for disorder in the system the boson energies 
$E_i^B=E^B+\delta E_i^B$ were randomly 
and uniformly distributed around 
level $E^{B}$ over the interval $\Delta E^B$. 
Physically this corresponds to fluctuating 
pairing potential \cite{Anderson06}, which in BF model 
arises in a very natural way.

Due to the above mentioned property of the 
model the magnitude 
of the local order parameter $<b_i>$ (not shown) obtained as a solution
 of equation (\ref{eq_BdG}) is  correlated with the 
position of the impurity provided we 
identify the dopant oxygens 
as e.g. those in Bi$_2$Sr$_2$CaCu$_2$O$_{8+\delta}$
with the pairing centres. From the same calculations it can also  be seen 
that both order parameters  $<b_i>$ and $\chi_i$ take on large values in the
same regions of space. The direct proportionality of both order parameters 
 is a general property of the model
\cite{Kostyrko}, which can also be inferred 
from the mean-field equation (\ref{ord_pars}).  

In figure (1) we show the effect of disorder  in bosonic 
(random $E_i^B$ levels - left panel)
and fermionic (random $V_i^{\text imp}$ - right panel) subsystems on the average values 
of the order parameters  $|<b_{i}>|$ and $|\chi_{i}|$.
\begin{figure}
\centerline{\epsfig{file=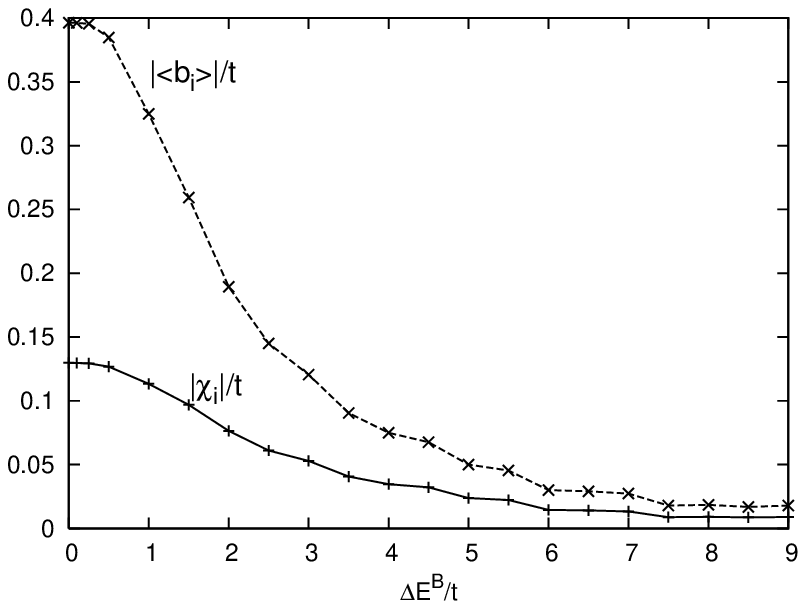,width=6cm} 
\epsfig{file=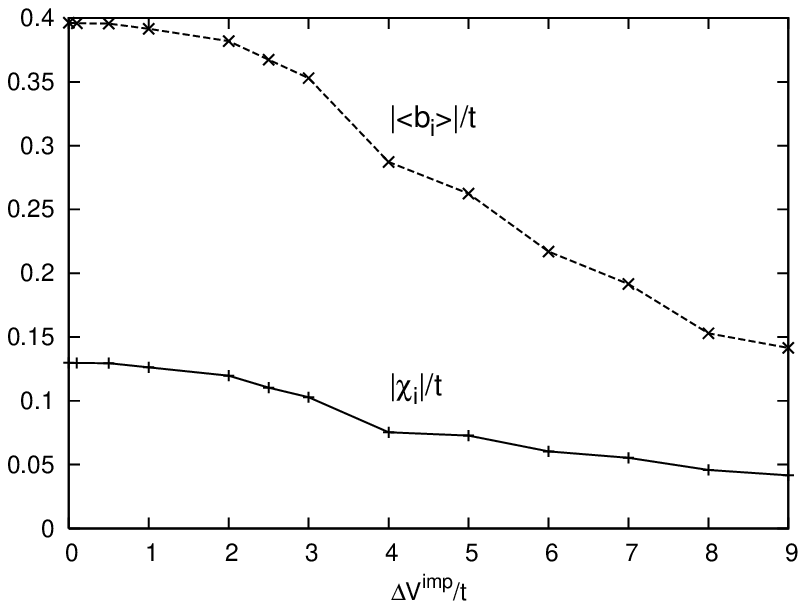,width=6cm}}
 \vspace{-4.0cm} 
\hspace{2.5cm}
\epsfig{file=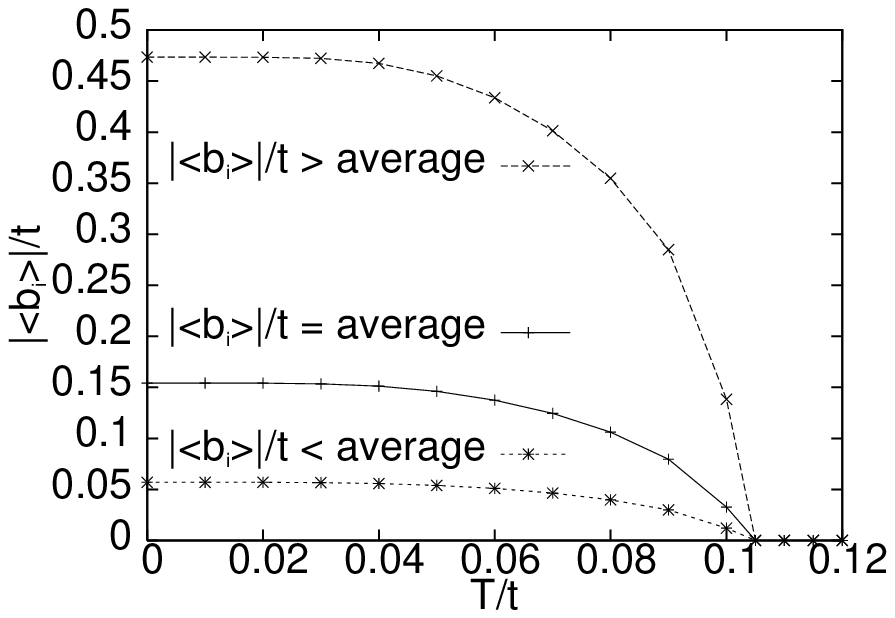,width=3.2cm}
\vspace{2.50cm} 
\caption{ Average values of $|<b_{i}>|$ and $|\chi_{i}|$ as 
a function of disorder  in the boson 
(left panel) and fermion (right panel) subsystem.
Inset to left panel: The temperature dependence of  
of average $|<b_{i}>|$ compared with similar dependence at
the sites at which the value of the gap is larger/smaller than average. 
The results were obtained for $t_{1}=t=1.0$,
 $t_{2}=-0.3t$, $\mu=0.0t$, $g=0.8t$ $E_B=-0.2t$.}
\end{figure} 
In this figure the clean system (characterised by $E_B=-0.2t$) 
would correspond to the 
experimental situation with all sites contributing
to boson-fermion scattering. With increase of $\Delta E^{B}$ 
the number of sites at which this scattering takes place
diminishes and the order parameter 
is strongly reduced.
 
It is interesting to note that despite the d-wave 
character of superconductivity the disorder acting on the
fermionic subsystem relatively weakly affects both order parameters
(figure (1), right panel).
More severe changes of both  $|<b_{i}>|$ and $|\chi_{i}|$ are observed for
very large disorder.
The results presented in inset to figure (1) 
show that the whole system 
is characterised by the unique transition temperature.
This result is correct from thermodynamical point of view.
At the same time it indicates that for the system studied here 
there exist a coupling
between regions of large and small pairing potential, preventing the
appearance of isolated regions with different values of the
transition temperatures, even though the gaps near $T_c$ do vary a lot 
as is also visible from figure (2) - right panel. 
Indeed rough 
estimation of the coherence length leads to values of order 1/3 to 1/2 
of the system size. This shows that all domains are coupled.
\begin{figure}
\begin{center}
\includegraphics[width=6.2cm]{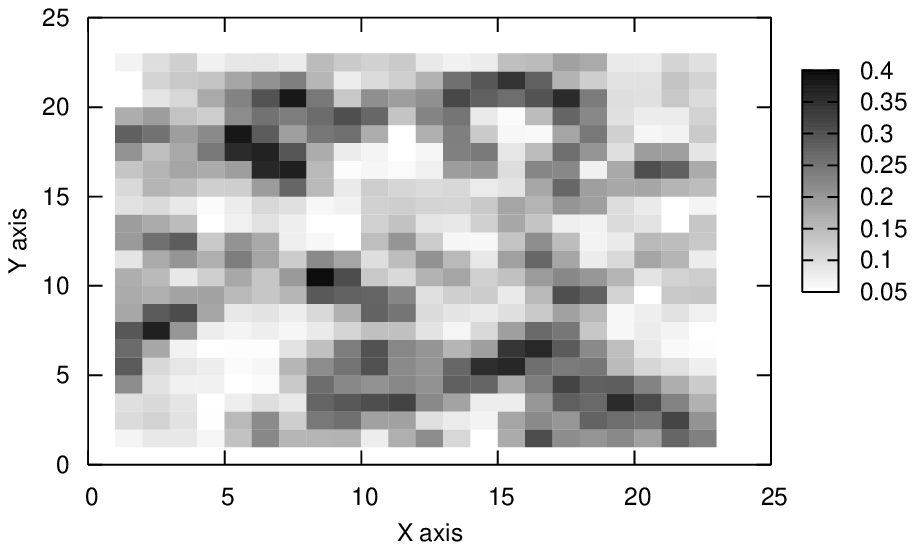}
\includegraphics[width=6.2cm]{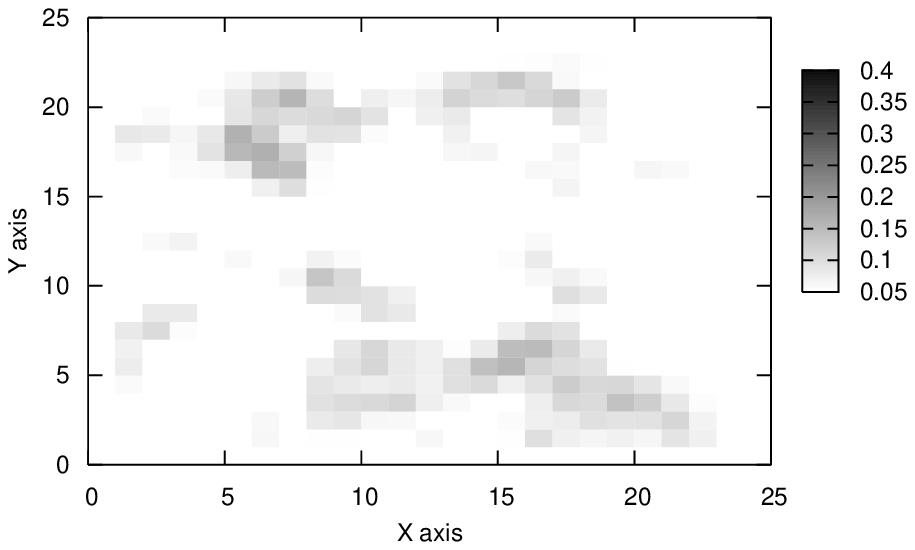}
\caption{Maps of amplitude of bosonic order parameter 
$|<b_{i}>|$ for different values of temperature: 
$T=0.0t$ (left panel), $T=0.1t\approx 0.9T_c$ (right panel). 
Other parameters the same as in figure (1)}
\end{center}
\end{figure}
In figure (2) we plot  the 
maps of bosonic order parameter $|<b_i>|$  
 of our 23$\times$23 large system at different temperatures.
They are shown in gray scale for two values of temperature -- 
 $T<<T_c$ (left panel)
and for $T\approx 0.9T_c$ (right panel). At temperatures
close to $T_c$ the regions of large gap seem to be 
surrounded  by the non-superconducting material. 
Close inspection shows that the whole system is 
superconducting but most of it is characterised
by very low value of the gap, while large gap persists
in few regions of the sample. We do not address the pseudogap
problem here, as our mean-field approach is not suitable
in that region of parameter space.

In conclusion, we studied the properties of
disordered  boson-fermion model allowing for two different 
disorder sources. The randomness in 
boson energies strongly degrades superconductivity, while 
 disorder acting on electron subsystem $V_i^{imp}$ has 
much weaker effect on superconducting state.
The local values of fermionic order parameter positively 
correlate with positions of "pairing impurities" characterised
by small values of $E_B$. Even strongly disordered system 
possesses single superconducting transition temperature despite
large fluctuations of the gap magnitude.\\
{\bf Acknowledgements:} This work has been partially 
supported by the grant no. N N202 1878 33. 

\end{document}